\documentclass[%
 aip,
 amsmath,amssymb,
 reprint,%
jcp]{revtex4-1}

\usepackage{graphicx}
\usepackage{dcolumn}
\usepackage{bm}
\usepackage[version=3]{mhchem}
\usepackage[utf8]{inputenc}
\usepackage[T1]{fontenc}
\usepackage{mathptmx}
\usepackage{etoolbox}
\usepackage{multirow}
\usepackage{booktabs}

\makeatletter
\def\@email#1#2{%
 \endgroup
 \patchcmd{\titleblock@produce}
  {\frontmatter@RRAPformat}
  {\frontmatter@RRAPformat{\produce@RRAP{*#1\href{mailto:#2}{#2}}}\frontmatter@RRAPformat}
  {}{}
}%
\makeatother
\begin{document}

\preprint{AIP/123-QED}

\title[Distinguishing homolytic versus heterolytic bond dissociation of phenyl sulfonium cations with localized active space methods]{Distinguishing homolytic versus heterolytic bond dissociation of phenyl sulfonium cations with localized active space methods}
\author{Qiaohong Wang} \affiliation{Pritzker School of Molecular Engineering, University of Chicago, Chicago, IL, 60637, USA.}
\author{Valay Agarawal}
\author{Matthew R. Hermes}
\affiliation{Department of Chemistry, Chicago Center for Theoretical Chemistry, University of Chicago, Chicago, IL 60637, USA.}
\author{Mario Motta}
\affiliation{IBM Quantum, IBM T. J. Watson Research Center, Yorktown Heights, NY 1059}
\author{Julia E. Rice}
\author{Gavin O. Jones}
\affiliation
{IBM Quantum, IBM Research-Almaden, San Jose, CA, 95120, USA.}
\author{Laura Gagliardi}
\email{lgagliardi@uchicago.edu}
\affiliation{Pritzker School of Molecular Engineering, University of Chicago, Chicago, IL, 60637, USA.}
\affiliation{Department of Chemistry, Chicago Center for Theoretical Chemistry, University of Chicago, Chicago, IL 60637, USA.}
\affiliation{James Franck Institute, University of Chicago, Chicago, IL 60637, USA.}
\affiliation{Argonne National Laboratory
9700 S. Cass Avenue
Lemont, IL 60439}
\date{\today}

\begin{abstract}
Modeling chemical reactions with quantum chemical methods is challenging when the electronic structure varies significantly throughout the reaction, as well as when electronic excited states are involved. Multireference methods such as complete active space self-consistent field (CASSCF) can handle these multiconfigurational situations. However, even if the size of needed active space is affordable, in many cases the active space does not change consistently from reactant to product, causing discontinuities in the potential energy surface. The localized active space SCF (LASSCF) is a cheaper alternative to CASSCF for strongly correlated systems with weakly correlated fragments. The method is used for the first time to study a chemical reaction, namely the bond dissociation of a mono-, di-, and triphenylsulfonium cation. LASSCF calculations generate smooth potential energy scans more easily than the corresponding, more computationally expensive, CASSCF calculations, while predicting similar bond dissociation energies. Our calculations suggest a homolytic bond cleavage for di- and triphenylsulfonium, and a heterolytic pathway for monophenylsulfonium. 
\end{abstract}

\maketitle

\section{\label{sec:1}Introduction}
Triphenylsulfonium (TPS) salts have been widely used for decades as photoacid generators (PAG) for lithographic processes in the semiconductor industry.\cite{tps_intro, tps_intro2}  The functionality of TPS salts comes from light-induced photolysis. When TPS salts are exposed to light, the C–S bond in the triphenylsulfonium cation (\ce{Ph_{3}S^{+}}) breaks, forming reactive intermediates that can induce reactions, such as acid-catalyzed cleavage or cationic polymerization that can change the solubility of substrates used in microelectronic fabrication.\cite{tps_function1,tps_function2,tps_function3,tps_function4,tps_function5} Experiments have suggested two different pathways of the dissociation of TPS salts, the homolytic bond-cleavage mechanism \cite{homo1,homo2,homo3}, the heterolytic mechanism \cite{hetero} and new decomposition pathways that include both \cite{both1,both2}. Time-dependent density-functional theory (TDDFT) calculations have been performed to calculate the absorption spectra of TPS salts in the geometry of the ground state \cite{DFT_tps}, but theoretical calculations on the photochemical decomposition of the TPS mechanism remain limited. Ohmori et al. \cite{ab_initio_tps} performed  configuration interaction singles (CIS) calculations for TPS and suggested a heterolytic pathway of the ground state until 3.6 \(\text{\AA}\) and an extrapolated homolytic product explained by an exchange mechanism of the ground state with the lowest singlet excited state. The results offer a limited explanation of the pathway since the CIS calculations were not carried past 3.6 \(\text{\AA}\), and whether such an exchange mechanism occurs remains unknown. To understand the ground-state dissociation completely, multiple electronic configurations may be required to model bond cleavage, especially in cases of the formation of radicals. To computationally describe such processes from equilibrium to dissociation, one often requires a multiconfigurational wave function.
The complete active space self-consistent field (CASSCF) \cite{casscf} method is the paradigmatic multireference method. In CASSCF, all electronic configurations generated by the active electrons occupying the active orbitals in all possible ways are included in the wave function. The number of configuration state functions (CSFs) grows exponentially with the size of the active space and limits the practicality of CASSCF to at most a (20e,20o) active space, even with parallelization.\cite{parallelmcscf}  In addition, many configurations have small CI amplitudes, making the CI expansion inefficiently large. There have been various efforts to reduce the cost of CASSCF, for example, the generalized active space self-consistent field (GASSCF),\cite{gasscf1,gasscf2,gasscf3,gasscf4} restricted active space self-consistent field (RASSCF),\cite{rasscf,rasscf2} quasi-CASSCF(QCASSCF),\cite{quasi} etc. Another strategy is to identify and divide the active space into separable fragments, exemplified by the localized active space self-consistent field (LASSCF) method,\cite{LAS,vLASSCF} also known as the cluster mean-field (cMF) method.\cite{Jimenez-Hoyos2015} LASSCF factorizes the active-space wave function into a single anti-symmetrized product of smaller active-space wave functions, under the assumption that such active spaces are relatively weakly interacting. The computational cost of determining the active-space CI vectors in LASSCF is thus linearly scaling with system size given fixed fragment size, and the cost scaling of the orbital optimization is the same as Hartree--Fock (HF).

CASSCF is a variational method in which the active orbitals and the CI coefficients are chosen to minimize the total energy. Therefore, if at a particular point on the potential energy curve of a certain reaction, the electrons in a given set of orbitals are not strongly correlated, it may qualitatively ``rotate those orbitals out'' to access more dynamical electron correlation elsewhere, even if the original orbitals \emph{were} strongly correlated at another point on the potential energy curve. Such orbital swapping may furthermore indirectly change the qualitative nature of the wave function along a reaction pathway (e.g., closed-shell to open-shell), making the description of chemical reactions challenging. This is one of the reasons why CASSCF is not easy to use for chemical reactivity, together with its combinatorial scaling. There have been attempts to automatize CAS-based orbitals for chemical reactions \cite{Stein_Reiher_2017,Bensberg,Wardzala_king} and this is an ongoing field of research 

In this paper, we study the dissociation of a single C-S bond in the monophenylsulfonium (\ce{PhSH2+}, MPS), the diphenylsulfonium (\ce{Ph2H2S+}, DPS) and the triphenylsulfonium (\ce{Ph3S+}, TPS) cations in the gas phase. Our results show that MPS in the ground electronic state dissociates heterolytically, whereas DPS and TPS dissociate homolytically. They also show, in the case of MPS, that the constraints on the LASSCF wave function imposed by the user-selected number of orbitals and electrons in each active-space fragment can prevent optimization to undesired local-minima wave functions, a problem to which CASSCF is more vulnerable. This feature helps to avoid some of the trial and error typical of CASSCF, making LASSCF a more intuitive and automated method, besides being computationally cheaper. Finally to get quantitative dissociation energies we complement the active space-based calculations with post-SCF treatments like perturbation theory and pair-density functional theory.

\section{\label{sec:2}Theoretical background and computational methods}
The LASSCF energy is optimized variationally by minimizing \cite{LAS,vLASSCF} 
\begin{equation}
    E_{\mathrm{LAS}}=\langle\mathrm{LAS}|\hat H| {\mathrm{LAS}}\rangle
\end{equation}
with a wave function ansatz 
\begin{equation}
|\mathrm{LAS}\rangle = \bigwedge_{K} (\Psi_{A_k}) \wedge \Phi_D
\end{equation}
where $\hat H$ is the molecular Hamiltonian, $\Psi_{A_{K}}$ is a general many-body wave function to describe electrons occupying active orbitals of the $A_{K}$-th fragment or ``active subspace,'' and $\Phi_{D} $ is a single determinant spanning the complement of the complete active space. The wedge operator $\bigwedge$  indicates that the active space wave function is an antisymmetrized product of $K$ fragment wave functions.\\
\begin{figure}
\includegraphics[scale = 0.4]{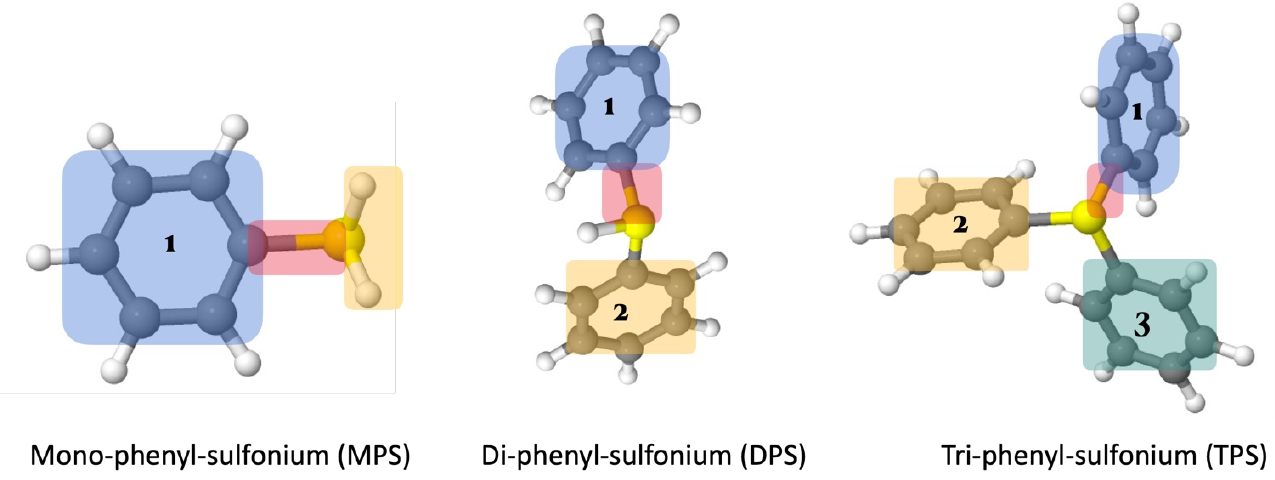}
\caption{\label{fig:3systems}Mono-, Di-, and Triphenylsulfonium systems with their fragmentations. The gray spheres are carbon atoms, the yellow sphere is a sulfur atom, and the white spheres are hydrogen atoms. The translucent blue, red, yellow and green boxes indicate the atoms associated with the three or four LASSCF fragments.}
\end{figure}
The MPS, DPS and TPS and their fragmentations are depicted in Figure \ref{fig:3systems}. MPS and DPS geometries are optimized with MP2 and DFT-B3LYP,\cite{BP86,B3LYP} and the TPS geometry is optimized with DFT-B3LYP. We observe identical bond angles and bond lengths from MP2 and DFT-B3LYP for both MPS and DPS. The dissociated geometries are generated by elongating the C-S bond (i.e., the potential energy scans discussed in the following are not geometrically relaxed). All three compounds have a singlet ground state and an overall (+1) charge. The process of determining chemically meaningful active spaces is described as follows. First, the most important LAS fragment is the one containing the C-S bond that dissociates. Then, we consider the phenyl that shares the carbon with the C-S bond to be the second most important (hereafter phenyl 1). Lastly, we consider fragments that remain with the sulfur atom at dissociation and their corresponding active spaces. The active space (AS) of MPS includes 12 electrons and 12 orbitals (12e,12o), which encompasses the carbon-sulfur (C-S) bond (2e,2o), the $\pi$ bonds of phenyl 1 that connect to the C-S bond (6e,6o) and the two sulfur-hydrogen (S-H) bonds (4e,4o) as shown in Figure \ref{fig:3systems}. For DPS, we considered three possible active spaces that encompass the C-S bond (2e,2o), all six $\pi$ orbitals of phenyl 1 (6e,6o), and the (0e,0o), (2e,2o) and (4e,4o) AS respectively, for the remaining phenyl (phenyl 2). We did not include the S-H bond in the AS because the phenyl 2 $\pi$ bonds are more likely to be important. For TPS, we explore ASs encompassing the C-S bond (2e,2o), the $\pi$ orbitals of phenyl 1 (6e, 6o), with (0e,0o), (2e,2o), and (4e,4o) for each of the remaining phenyl’s $\pi$ bonds (phenyl 2 and 3). 

We report dissociation energies obtained with CASSCF, LASSCF, and CASCI with the LASSCF orbitals to capture the missing inter-fragment correlations from LASSCF (hereafter CASCI). To include dynamic correlation, we performed n-electron valence state perturbation theory NEVPT2,\cite{nvept2_1,nevpt2_2} calculations using CASSCF and CASCI reference wave functions. We also perform multiconfiguration pair density functional theory (MC-PDFT) calculations, with the translated Perdew-Burke-Ernzerhof (tPBE) functional, hereafter referred to as CAS-PDFT,\cite{mcpdft1,mcpdft2,mcpdft3} and LAS-PDFT,\cite{laspdft} using the CASSCF and LASSCF reference wave functions, respectively. Kohn-Sham DFT calculations with the B3LYP-D3BJ \cite{BP86,B3LYP}, TPSSh-D3BJ\cite{TPSSH}, BP86-D3BJ\cite{BP86}, M062X\cite{M062X}, and MN15\cite{MN15} functionals were also carried out. All DFT calculations are performed with the def2tzvp basis set \cite{def2tzvp} using the Gaussian package version 16A01\cite{g16}, and the multireference calculations with the cc-pVDZ basis set \cite{ccpvdz} and PySCF \cite{Pyscf} with the mrh extension\cite{mrh}, with the exception that we used OpenMolcas \cite{OpenMolcas} for CASSCF calculations involving MPS. 

\section{\label{sec:3}Results and discussion}
\subsection{\label{sec:3.1}Optimized wave function at equilibrium and dissociation}
\begin{figure}
\includegraphics[scale = 0.4]{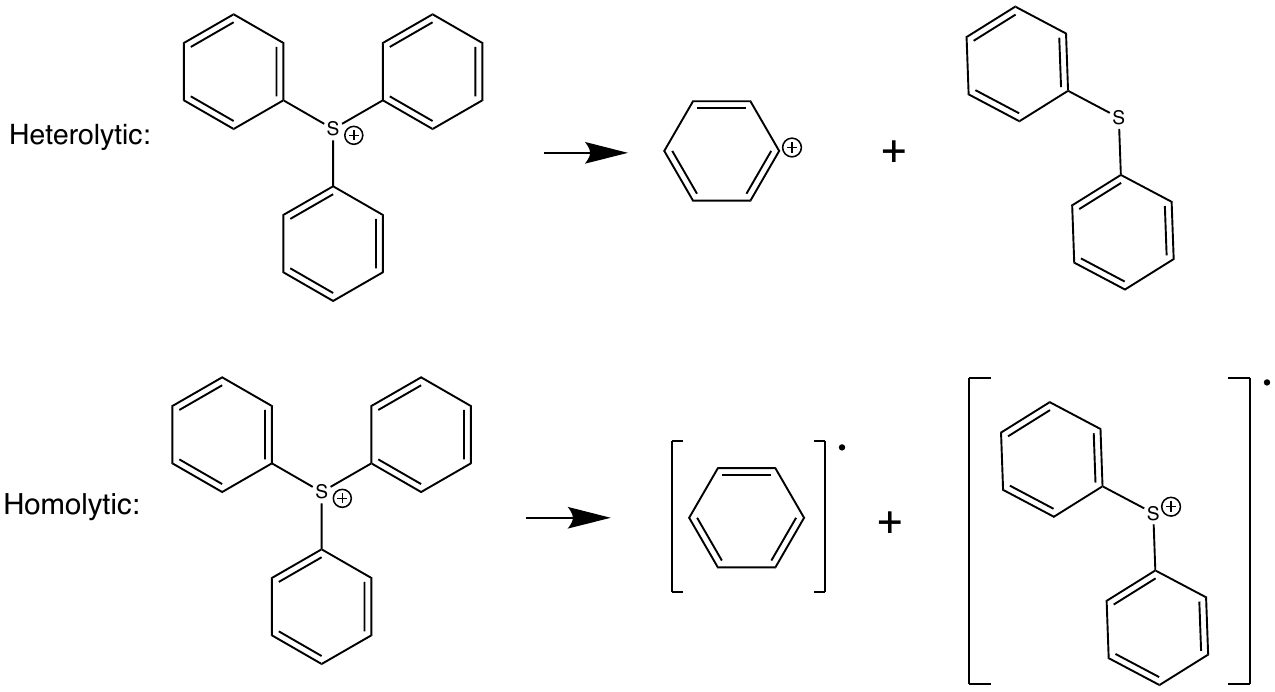}
\caption{\label{fig:tps_pathway}Possible dissociation pathways for the triphenylsulfonium cation. The heterolytic pathway will generate a phenyl cation and di-phenyl sulfonium. The homolytic pathway will generate a phenyl radical and di-phenyl sulfonium radical.}
\end{figure}

The bond dissociation can occur either heterolytically or homolytically, as illustrated pictorially for TPS in Figure \ref{fig:tps_pathway}.
To determine the dissociation pathway, we first perform calculations at the equilibrium geometry and at dissociation (that is, the C-S bond at 6 \text{\AA} without additional optimization) as described below.

\begin{figure*}
\includegraphics[scale = 0.75]{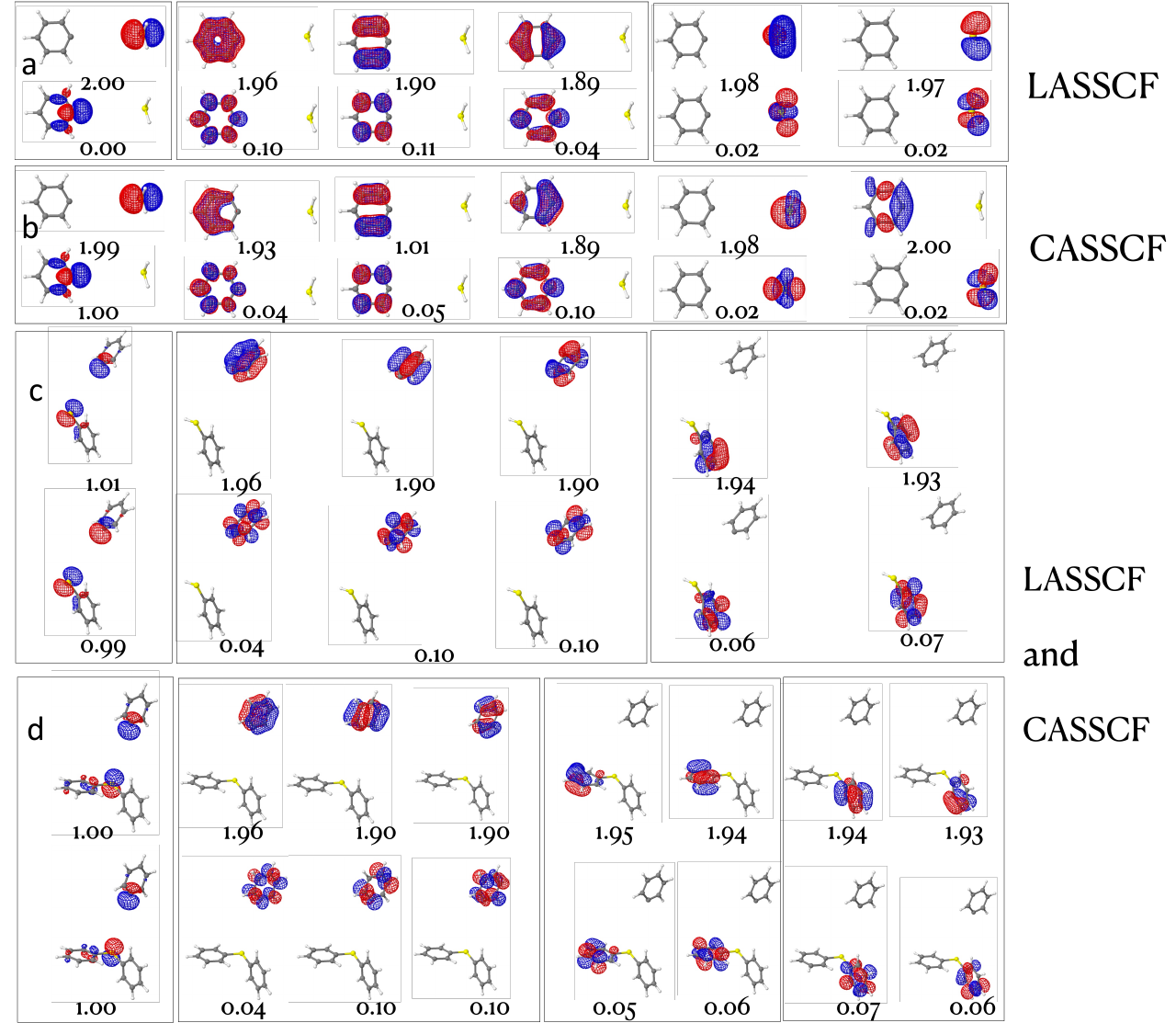}
\caption{\label{fig:dissociated}a. Dissociated product of MPS with LASSCF, the natural orbital occupancy number of the C-S bonding orbital 2.00 and anti-bonding orbital 0.00 suggest a heterolytic product. b. Dissociated product of MPS with CASSCF, the natural orbital occupancy numbers of the C-S bonding orbital 1.99 suggest a heterolytic product. Note the hopping of electron from $\pi$ system to C-S bonding/anti-bonding fragment. c. Dissociated product of DPS, the natural orbital occupancy numbers 1.01 and 0.99 suggest homolytic results.  d. Dissociated product of TPS, the natural orbital occupancy numbers 1.00 and 1.00 suggest homolytic results.  }
\end{figure*}
At equilibrium, we used the atomic valence active space (AVAS) method,\cite{AVAS} operating on the closed-shell spin-singlet restricted Hartree--Fock (RHF) wave function, to generate the valence orbitals (i.e. linear combinations of C ‘2$p$’ and S ‘3$p$’), used as an initial guess for the subsequent CASSCF and LASSCF calculations. At dissociation, we initialized the CASSCF and LASSCF orbitals in two separate ways, designed to target a closed-shell dissociated product in one case and an open-shell dissociated product in the other. We then retained the result with the lower total energy. In the first case, we followed the same procedure as for the equilibrium geometry: closed-shell spin-singlet RHF followed by AVAS; we call this ``singlet guess.''  In the second case, we applied AVAS to a spin-triplet restricted open-shell Hartree--Fock (ROHF) wave function, followed by spin-triplet CASSCF or LASSCF on the resulting set of orbitals, followed by spin-singlet CASSCF or LASSCF using the optimized triplet orbitals as the starting point; we call this ``triplet guess.''

The use of the CASSCF and LASSCF methods poses different challenges, which are typical when active space-based methods are employed and are the reason why these methods are not considered black-box methods. The first challenge is to choose a meaningful active space along the reaction, which in the CASSCF case involves trial and error. For LASSCF, the user needs to choose the fragments and their relevant bonding and anti-bonding orbitals. The next challenge is to converge the CASSCF and LASSCF wave functions starting from the initial guess. This is usually a challenge for CASSCF. The extra information needed from LASSCF turns out to be advantageous since it simplifies the active space and exerts better control of meaningful orbitals, resulting in faster convergence and more precise active space depiction in accordance with user preference. One example of the issues that one can encounter in CASSCF calculations involves MPS at both equilibrium and dissociation. The singlet guess routine described above did not give a result that contains the desired orbitals in the active space. Therefore, for MPS, the singlet-guess CASSCF results were instead obtained by initializing with optimized orbitals from a state-averaged CASSCF with four roots (SA(4)-CASSCF) calculation carried out in OpenMolcas, at both the equilibrium and dissociated geometries.

\begin{table}[b!]
\renewcommand{\arraystretch}{1.3}
\setlength{\tabcolsep}{5 pt}
\caption{\label{tab:3comp_table}Dissociation energy (in eV) comparison for MPS, DPS, TPS with their largest active spaces, calculated as the difference between equilibrium and dissociated geometry. Singlet guess led to wave functions corresponding to heterolytic dissociation, and triplet guess led to wave functions corresponding to homolytic dissociation.}
\begin{tabular}{|l|l|l|l|}
\hline
System & Method & Singlet guess & Triplet guess \\ \hline
\multirow{4}{*}{\begin{tabular}[c]{@{}l@{}}MPS\\ (12e,12o)\end{tabular}} & CASSCF & 3.4\textsuperscript{\emph{b}} & 4.1 \\ 
 & LASSCF & 3.7 & 4.0 \\ 
 & NEVPT2(CASSCF) & 3.8 & 4.4 \\  
 & NEVPT2(CASCI\textsuperscript{\emph{a}}) & 4.0 & 4.4 \\ \hline
\multirow{4}{*}{\begin{tabular}[c]{@{}l@{}}DPS\\ (12e,12o)\end{tabular}} & CASSCF & 3.9 & 3.8 \\ 
 & LASSCF & 4.6 & 3.8 \\ 
 & NEVPT2(CASSCF) & 5.0 &3.9 \\  
 & NEVPT2(CASCI\textsuperscript{\emph{a}}) & 5.0 & 4.1 \\ \hline
\multirow{4}{*}{\begin{tabular}[c]{@{}l@{}}TPS\\ (16e,16o)\end{tabular}} & CASSCF & 4.4 & 3.1 \\ 
 & LASSCF & 5.0 & 3.3 \\ 
 & NEVPT2(CASSCF) & 5.4 & 3.5 \\  
 & NEVPT2(CASCI\textsuperscript{\emph{a}}) & 5.5 & 3.7 \\ \hline
\end{tabular}

\textsuperscript{\emph{a}} LAS orbitals \\
\textsuperscript{\emph{b}} Upper bound, see Sec.\ \ref{sec:3.2}
\end{table}
The optimized wave functions at dissociation for all three systems are shown in Figure \ref{fig:dissociated}. In the case of MPS, it results from the singlet (i.e., OpenMolcas SA(4)-CASSCF) guess, while DPS and TPS results from the triplet guess. The full results of the singlet-initialized and triplet-initialized calculations are shown in Figure S1 of the Supporting Information. A comparison of dissociation energies obtained with the different guesses is listed in Table \ref{tab:3comp_table}. We can identify the dissociation pathway by examining the product and its characteristic orbital occupancies at dissociation. For example, the C-S bonding and antibonding orbitals in Figure \ref{fig:dissociated}\textbf{a} show occupancies of 2.0 and 0.0, which is the characteristic of a heterolytic product, while the occupancy numbers of 1.0 and 1.0 of the C-S bonding and antibonding orbitals for \textbf{c} and \textbf{d} indicate a homolytic product. The natural orbital occupancy of the optimized wave function at the dissociation suggests a heterolytic pathway for MPS, and a homolytic pathway for DPS and TPS.

We further confirm that our LASSCF wave functions all model nondegenerate ground states by using both singlet- and triplet-initialized LASSCF orbitals as a starting point for CASCI calculations (see Tables S1, S2, S3, S4, S5 and S6 in the Supporting Information). We characterize each state and present their energies to confirm that our ground state solution is heterolytic for MPS and homolytic for DPS and TPS, and nondegenerate in all cases (the smallest observed singlet excitation energy in any case was 1.02 eV).

In contrast, the CASSCF (Figure \ref{fig:dissociated}\textbf{b}) and LASSCF (Figure \ref{fig:dissociated}\textbf{a}) wave functions of the dissociated product of MPS qualitatively differ. Although both methods predict heterolytic dissociation, CASSCF predicts an open-shell state in which an electron from the phenyl $\pi$ system has hopped into the empty carbon 2$p$ atomic orbital that was formerly bonded to sulfur. Additionally, a qualitative ``orbital rotation'' occurred in the CASSCF calculation, in which one of the sulfur S-H bonds was substituted for a phenyl $\sigma$-system bond. On the other hand, the LASSCF calculation was constrained in such a way that the $\pi\to2p$ state, which would have corresponded to 3 electrons in the first fragment and 5 electrons in the second, is forbidden, and all orbitals remained qualitatively similar to the guess orbitals. We address the question of whether CASSCF calculation predicted the lowest product ground state, or whether it became trapped in a local minimum, by seeking a smooth potential energy scan that connects the optimized wave function from equilibrium to dissociation. Using both the equilibrium wave function (forward-scanning) and the optimized wave function at dissociation (backward-scanning) as a start, we performed potential energy scans for all three systems.

\subsection{\label{sec:3.2}Dissociation pathways}

\begin{figure*}
\includegraphics[scale = 0.5]{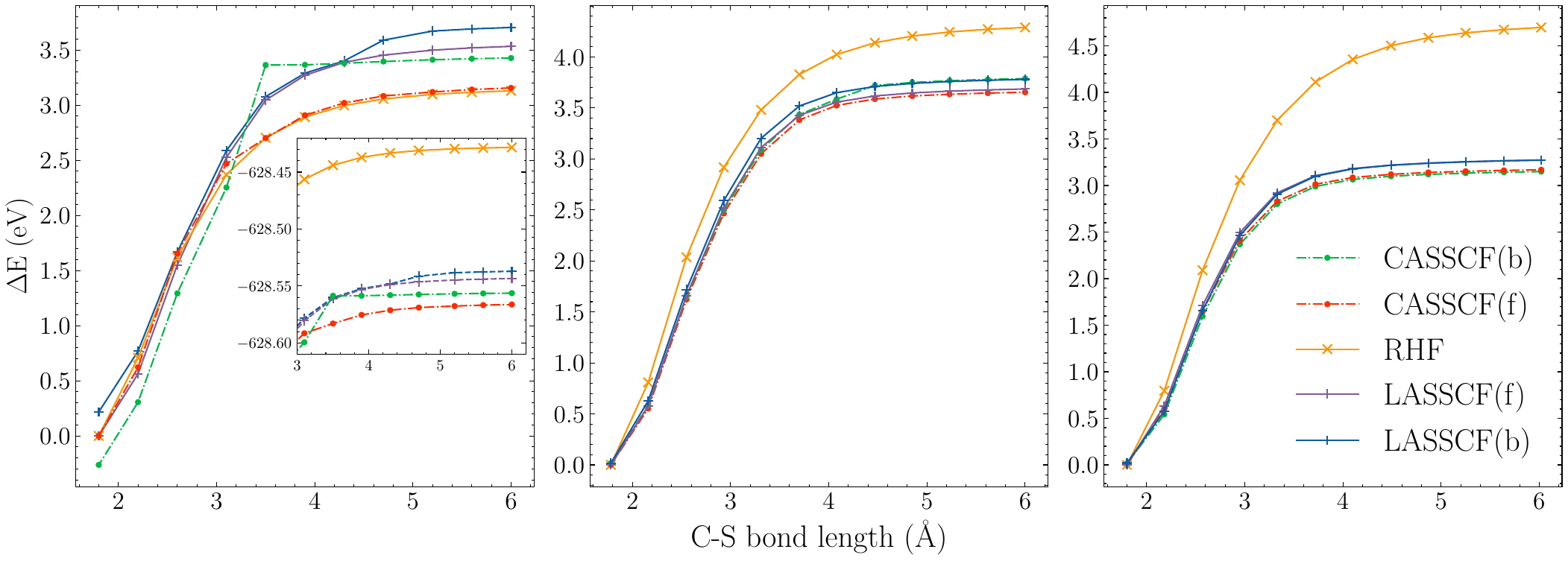}
\caption{\label{fig:pe_scan}\textit{Left}: Ground state, heterolytic dissociation potential energy scan of MPS cation, inset shows the absolute energies. \textit{Middle}: Ground state, homolytic dissociation potential energy scan for DPS cation. \textit{Right}: Ground state, homolytic dissociation potential energy scan for TPS cation. Energy is plotted relative to that at the equilibrium geometry in all cases, and “(b)" indicates scan initiated from dissociation, while “(f)" indicates scan initialized from the equilibrium geometry.}
\end{figure*}

By exploring the natural orbital occupancies along the potential energy scans following the backward scan, we confirm a heterolytic pathway for the ground-state MPS cation and homolytic pathways for both the ground-state DPS cation and the TPS cation.  The potential energy curves of the MPS cation (12e, 12o), the DPS cation (12e, 12o), and the TPS cation (16e, 16o) are shown in Figure \ref{fig:pe_scan}. 

First, both CASSCF and LASSCF describe a smooth homolytic dissociation for the DPS cation and the TPS cation. Forward and backward potential energy scans with LASSCF and CASSCF are in close agreement, as indicated in the two right panels of Figure \ref{fig:pe_scan}. See Section 3, Figures S2 and S3 in the Supporting Information for a more complete examination of the LASSCF orbitals and occupation numbers along the dissociation curve. 

\begin{figure*}
\includegraphics[scale = 0.65]{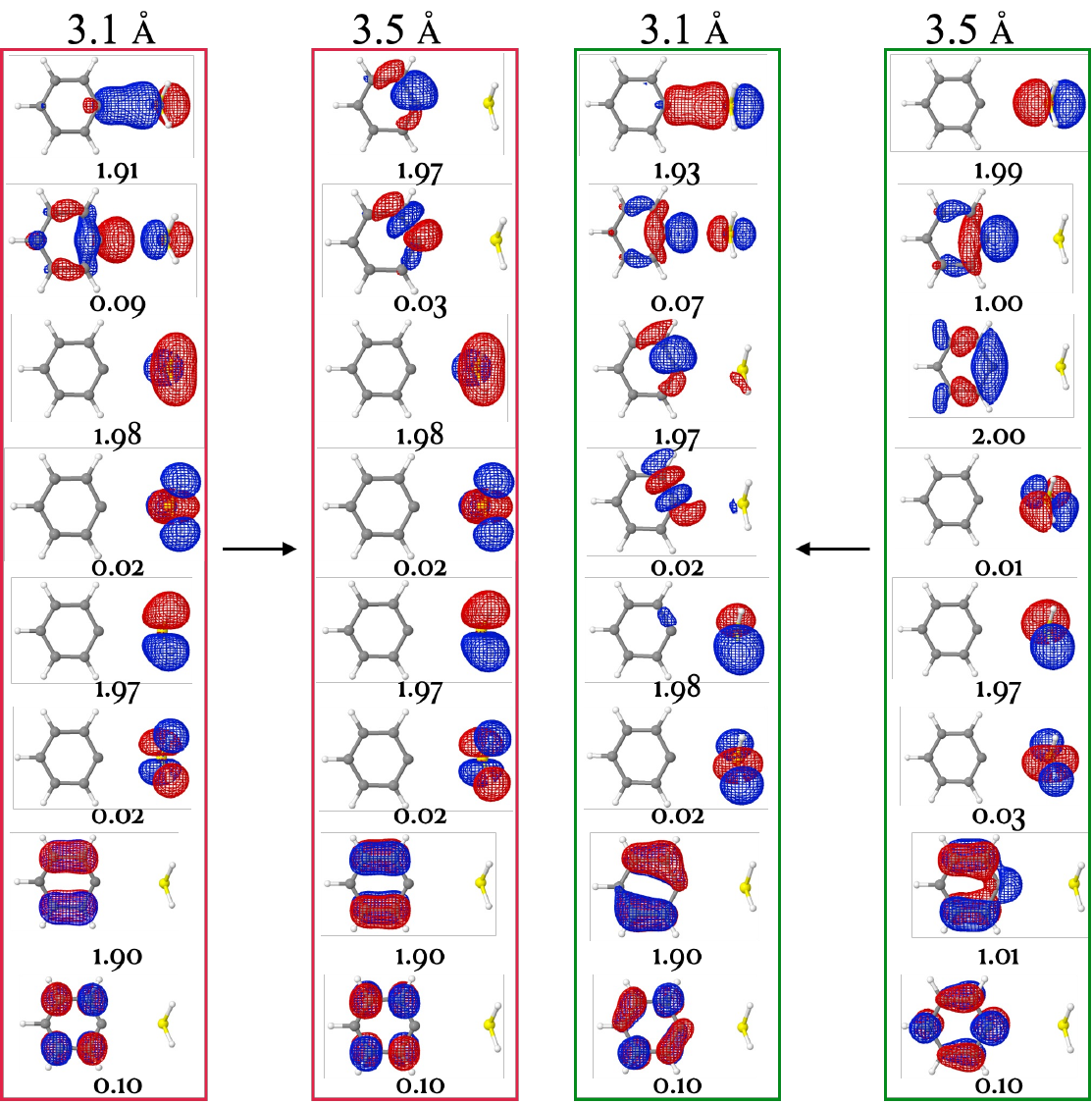}
\caption{\label{fig:mps_cas}\textit\textit{Left}: 8 orbitals from CASSCF(f) (12,12) potential energy scans at different bond lengths.  \textit{Right}: 8 orbitals from CASSCF(b) (12,12) potential energy scans at different bond lengths. Note the change in occupation at of 2nd and 7th orbitals at 3.5$\AA$.  }
\end{figure*}

Second, the single-reference nature of the MPS heterolytic dissociation appears challenging for CASSCF, while LASSCF shows significantly more control over the wave function along the scan. We first observe that LASSCF(f) and LASSCF(b) remain qualitatively similar and relatively smooth throughout the scan, while CASSCF(b) has a visible discontinuity starting at 3.1  \(\text{\AA}\), and CASSCF(b) and CASSCF(f)  differ considerably. Despite the smoothness of the CASSCF forward curve, upon further inspection of the orbitals, as shown in the red boxes in Figure \ref{fig:mps_cas}, we identified a pair of carbon-carbon (C-C) $\sigma$ bonds that are swapped into the previously defined active space at 3.5 \text{\AA}, replacing the C-S bond. The rest of the orbitals are consistent throughout the dissociation, and the C-C bonds remain in the active space until the dissociation limit.

\begin{figure*}
\includegraphics[scale = 0.65]{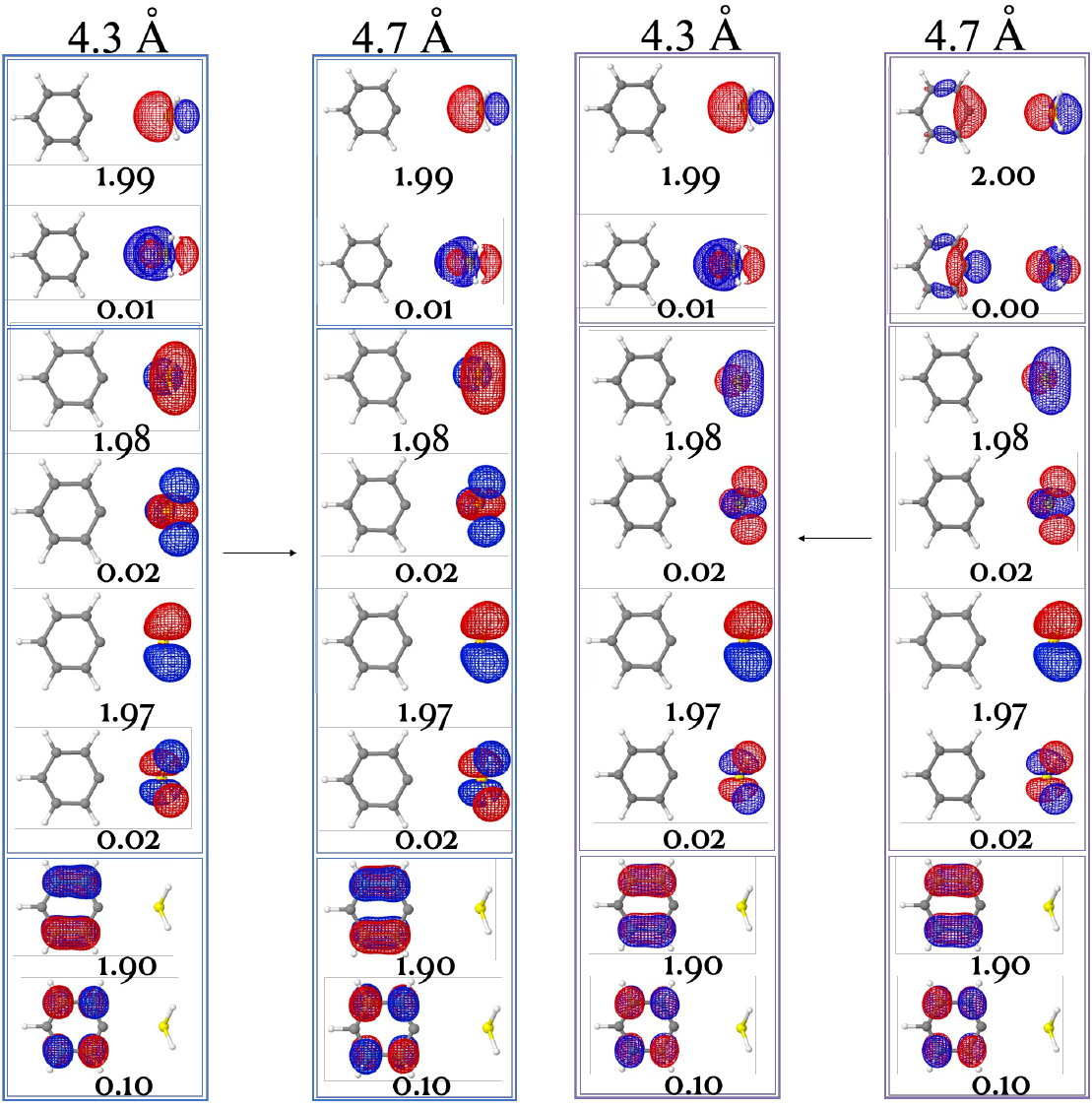}
\caption{\label{fig:mps_las}\textit{Left}: 8 orbitals from LASSCF(f) ((2,2),(6,6),(4,4)) potential energy scans at different bond lengths.  \textit{Right}: 8 orbitals of LASSCF(b) ((2,2),(4,4),(6,6)) potential energy scans at different bond lengths. The boxed orbitals indicate the corresponding fragments. }
\end{figure*}
The discontinuity of the CASSCF(b) curve, as depicted by the the green boxes in Figure \ref{fig:mps_cas} corresponds to a qualitative change from an open-shell to a lower-energy closed-shell wave function. It is clear that the initial CASSCF wave function of the dissociated product which initialized the backwards scan was a local minima characterized by a $\pi\to p$ excitation of the phenyl cation.

LASSCF, on the other hand, as expected, observes the user-imposed restriction of six electrons within the $\pi$ system and two electrons within the C--S bond-antibond system. It therefore generates both forward-scanning and backward-scanning curves more smoothly than CASSCF, although not perfectly smoothly. Figure \ref{fig:mps_las} presents the orbitals in the region of the slight bump in LASSCF(b) between 4.3 and 4.7 \text{\AA}. Clearly LASSCF is not entirely immune to ``orbital inconsistency,'' as the unoccupied C-S antibonding orbital is readily substituted for a high sulfur virtual orbital. However, this change does not qualitatively alter the wave function, and it still describes a heterolytic dissociation between the carbon atom and sulfur atom. These aforementioned orbital swapping artifacts also resulted in dissociation energy disagreements among the scans. For example, CASSCF(b) gives an equilibrium wave function that is lower in energy, indicated by the below zero point at equilibrium, while LASSCF(b) predicts a higher energy equilibrium wave function.

While one may argue that one can always improve the CASSCF scan, LASSCF inherently reduces the complexity of the problem. The price of LASSCF is that the user must divide the active space into fragments, but in practice active spaces are often intuitively constructed by users. 
Moreover, the LASSCF wave function has exponentially fewer degrees of freedom and is therefore much less likely to change character in an uncontrolled or undesired way, such as by an electron hopping from one fragment to another.
That said the LASSCF result is not expected to be as accurate as CASSCF in terms of relative energies because LASSCF is a drastic approximation to CASSCF. However, if the wave function is qualitatively correct, LASSCF can be a good starting point for post-SCF calculations. 

\subsection{\label{sec:3.3}Effects of dynamical electron correlation and advantage of multireference methods}
In the examples presented here, in addition to qualitatively describing physically meaningful orbitals along the dissociation, LASSCF also reports energies comparable to CASSCF. We now compared the active-space based dissociation energies with those obtained with single reference Kohn-Sham DFT, KS-DFT. 

We report the heterolytic dissociation energy of MPS in Table \ref{mps_table}. The LASSCF dissociation energy is computed as the difference between the two optimized wave functions discussed in Section 3.1. For CASSCF, we take the wave function at dissociation from the forward-scanning CASSCF curve as the dissociated product. We report the absolute energies of the aforementioned new CASSCF results in SI Table S8.
We further report the homolytic dissociation energy of DPS and TPS in Table \ref{dpstps}. The CASSCF and LASSCF dissociation energy is taken as the difference between the two optimized wave function we find in Section 3.1.
\begin{table}[b!]
\renewcommand{\arraystretch}{1.3}
\setlength{\tabcolsep}{7.5pt}
\caption{\label{mps_table}Comparison of the heterolytic dissociation energy of MPS cation by various methods. The dissociation energy of LASSCF and relevant methods are calculated as the energy difference between equilibrium and dissociated geometry. The dissociation energy for CASSCF reported here is from the forward scan.}
    \begin{tabular}{lc}
    \hline\hline
    \multicolumn{1}{c}{\bf{Method}}&
    \multicolumn{1}{c}{\bf{Heterolytic dis. energy (eV)}}\\
    \hline
    CASSCF(12,12) & 3.2  \\
    LASSCF((2,2)(4,4),(6,6)) &3.7  \\
    CASCI\textsuperscript{\emph{a}} & 3.9 \\
    NEVPT2(CASSCF)& 4.1 \\
    NEVPT2(CASCI\textsuperscript{\emph{a}}) &4.0\\
    LAS-PDFT\textsuperscript{\emph{*}}& 3.8\\
    CAS-PDFT\textsuperscript{\emph{*}}& 3.5\\
    \hline
    \end{tabular}
    
    \textsuperscript{\emph{a}} LAS orbitals;\textsuperscript{\emph{*}} tPBE functional

\end{table}

For MPS, the LASSCF dissociation energy is within 0.2 eV of CASCI and 0.3 eV of NEVPT2. CASSCF, on the other hand, differs by 0.5 eV from LASSCF, but the corresponding NEVPT2 result differs by only 0.1 eV from that of LASSCF. The LAS-PDFT value of 3.8 eV is similar to the NEVPT2 value, while the CAS-PDFT value is slightly lower at 3.5 eV. For DPS and TPS, the LASSCF energies are within 0.2 eV to the CASSCF value, and within 0 eV to 0.1 eV to its CASCI limit across all active spaces. We again observe close agreements among multireference methods and MC-PDFT calculations. We also note that the NEVPT2(CASCI) energies of the largest active space for both DPS and TPS increase only 0.1 eV compared to the smallest active space. In addition, the lower dissociation energy of TPS compared to DPS may be due to resonance effects in the SPh$_2$ radical that makes the radical products more stable, which can be observed as the slight delocalization of the singly-occupied sulfur 2$p$ orbital across the $\pi$ orbitals of one of the phenyl rings, which is not observed for DPS (see SI Section 1). Tables S7, S8, S9, and S10 in the SI include a more complete comparison of the absolute energies of equilibrium, heterolytic, and homolytic dissociated products in various methods.
\begin{table}[h!]
\renewcommand{\arraystretch}{1.3}
  \caption{Comparison of the homolytic dissociation energies for DPS and TPS (in eV).}
   \begin{tabular}{lcccccc}
\hline\hline
      & \multicolumn{3}{c}{DPS cation} & \multicolumn{3}{c}{TPS cation} \\
\cmidrule(lr){2-4}\cmidrule(lr){5-7}
Method & (8,8) & (10,10)  & (12,12) & (8,8) & (12,12) & (16,16)\\ \hline CASSCF & 3.6&3.9& 3.8&3.2&3.1&3.1\\
 LASSCF &3.6 &3.9& 3.8&3.1&3.1&3.3\\
 CASCI\textsuperscript{\emph{a}} &3.6 &3.9  &3.8 &3.2&3.1&3.2\\
 NEVPT2(CASSCF)&4.0&3.7&3.9&3.6&3.3&3.5\\
 NEVPT2(CASCI\textsuperscript{\emph{a}} )&4.0& 3.9& 4.1&3.6&3.6&3.7\\
 CAS-PDFT\textsuperscript{\emph{*}}&4.2&4.2 &4.2&3.6&3.5&3.6\\
 LAS-PDFT\textsuperscript{\emph{*}}&4.2& 4.5&4.5&3.6&3.6&3.8\\
\hline
\end{tabular}

\textsuperscript{\emph{a}} LAS orbitals;\textsuperscript{\emph{*}} tPBE functional
\label{dpstps}%
\end{table}%

The ground-state wave function of MPS is consistently qualitatively single-determinantal across the entire potential energy curve, and for such wave functions, single-reference methods such as KS-DFT are usually quantitatively accurate. However, KS-DFT results depend on the user's choice of exchange-correlation functional. To explore whether single-reference methods may be appropriate in this case, we perform KS-DFT calculations for equilibrium and dissociation. At equilibrium, and when targeting the heterolytically dissociated, closed-shell product, we use spin-restricted Kohn-Sham (RKS)-DFT. To target the homolytically dissociated, open-shell product, we use spin-unrestricted Kohn-Sham (UKS)-DFT calculations and a guess electron density corresponding to the desired numbers of electrons on each dissociated product. The dissociation energy is reported in Table \ref{dfttable} using the lower between the UKS and RKS total energies in the dissociated limit. See SI Table S11 for the absolute energies. We also report the absolute magnitude of spin density on each dissociated product in Table \ref{dfttable} (the spin density of the two dissociated products has the same magnitude but differs by a sign). Therefore, a value close to 1 corresponds to a homolytically dissociated, open-shell product, and a value close to 0 corresponds to a heterolytic dissociated, closed-shell product. However, UKS-DFT calculations produce ambiguous fractional numbers of spin density that are intermediate between 0.0 and 1.0. We therefore characterize a spin density from 0.7 to 1.0 to be homolytic, while 0.0 to 0.3 to be heterolytic. 

\begin{table*}
\caption{\label{dfttable}DFT results for MPS, DPS and TPS.  Dissociation is taken as the lower energy result between UKS-DFT and RKS-DFT}
\renewcommand{\arraystretch}{1.3}
\setlength{\tabcolsep}{9 pt}
\begin{tabular}{clcl}
\hline\hline
\multicolumn{1}{c}{Molecule}&
\multicolumn{1}{c}{Functional}&
\multicolumn{1}{l}{Dissociation energy (eV)} &
\multicolumn{1}{c}{Spin density and Character} \\ 
\hline
\multirow{5}{*}{MPS}
&B3LYP-d3bj& 3.3&0.36, N.A. \\
&TPSSh-d3bj& 3.1 & 0.37, N.A.  \\ 
&BP86-d3bj& 2.9 &0.38, N.A.\\
&M062X& 3.8 & 0.25, heterolytic\\
&MN15&3.7& 0.00, heterolytic\\
\hline
\multirow{5}{*}{DPS}
&B3LYP-d3bj&3.6&0.64, N.A.\\
&BP86-d3bj&3.3&0.59, N.A. \\
&TPSSh-d3bj&3.7&0.63, N.A. \\
&M062X&4.1&0.76, homolytic\\
&MN15&3.9 &0.66, N.A.\\
\hline
\multirow{5}{*}{TPS}
&B3LYP-d3bj & 3.5& 0.82, homolytic \\
&TPSSh-d3bj& 3.4 &0.78, homolytic \\
&BP86-d3bj & 3.3&0.71, homolytic  \\
&M062X &3.7 & 1.00, homolytic \\
&MN15 &3.7 &0.91, homolytic\\
\hline

\end{tabular}
\end{table*}

For MPS, UKS-DFT results provide a lower energy solution for 4 out of 5 functionals. Despite being lower in energy, this solution does not accurately depict the unpaired electrons configuration for the phenyl radical and SH$_2^{+}$ radical as indicated by spin density in Table \ref{dfttable}. The symmetry of the system is broken during the UKS-DFT calculation, leading to a fractional spin density that is less than 0.4 for all five functionals. As a result, DFT appears to predict neither a homolytic nor a heterolytic product for MPS dissociation, except for the MN15 and M062X functional, which predicts heterolytic dissociation. For DPS, all five UKS-DFT calculations lead to the lower dissociation energy, but only M062X predicts a spin density that can be unambiguously characterized, in this case as a homolytic dissociated product. For TPS, all five functionals predict a homolytic results with reasonable spin densities. 

NEVPT2 and PDFT calculations, on the other hand, confirmed the qualitative results of the LASSCF and CASSCF calculations: in no case did NEVPT2 or PDFT reverse the energy order of the homolytic and heterolytic dissociated products of MPS, DPS, or TPS (see Table 7, 9, and 10 in the SI respectively). This is an example of cases in which multireference methods can provide more accurate electronic structure information, although they are computationally more expensive than DFT.

\section{\label{sec:4}Conclusion}
Describing bond-breaking chemical reactions is a challenging and crucial task in computational chemistry. In the specific cases of the mono-, di- and triphenylsulfonium systems, we observe several advantages of using LASSCF to understand dissociation. First, LASSCF provides additional flexibility in choosing localized active orbitals to capture specific electronic configurations or states. This user-defined choice simplifies the active space problem by intuitively constructing chemically meaningful fragments, thereby avoids trial and error, 
which is often needed for performing CASSCF calculations.
The fragmentation of the active space provided by LASSCF is chemically driven, as shown by LASSCF reporting energetics that are comparable to those of CASSCF's for the systems studied here. Second, we note that LASSCF qualitatively evolves the same set of orbitals in both heterolytic and homolytic pathways. This feature is shown to be more beneficial in the heterolytic case, where CASSCF prioritizes less chemically meaningful orbitals. In addition, we show that our LASSCF's strength remains when the complexity of the systems increases, and the features that control the orbitals involved in the reaction accurately remain the same.

We want to emphasize here that LASSCF is a drastic approximation to CASSCF and the agreement between the two methods in terms of dissociation energies is specific to the systems investigate here. We also conclude that the MPS ground-state dissociation is heterolytic, while that of DPS and TPS is homolytic. Currently, excited state methodologies for LASSCF, namely LAS-state interaction (LASSI), are under development. The preliminary results of the ground-state dissociation we report here provide a solid foundation for further investigations of the excited states of TPS and their dissociation pathways. 

The present work has meaningful implications for research at the interface between electronic structure and quantum computation. In published literature, orbital optimization has been integrated in the workflow of variational quantum computing simulations as a way of enhancing the expressive power and optimization landscape of both physics-inspired \cite{sokolov2020quantum,zhao2023orbital} and hardware-efficient ansatzes \cite{moreno2023enhancing}, and as a compelling framework to compute response functions \cite{mizukami2020orbital}. However, orbital optimization is a delicate operation because active-space solutions yielded by quantum computing algorithms for near-term devices are imperfect due to algorithmic approximations (e.g. ansatz quality), decoherence, and shot noise \cite{de2023complete}.
The present work highlighted that LASSCF orbitals are more robust than CASSCF orbitals. This observation in turn suggests that LASSCF orbitals may be robust under the imperfections of active-space solutions mentioned above. LASSCF thus appears as a compelling candidate for a study where quantum computing methods for near-term devices are used in lieu of exact diagonalization as active-space solvers. Research in this direction is underway.

\begin{acknowledgments}
The authors thank Abhishek Mitra and Dr. Ruhee D'Cunha for their helpful discussion and feedback of this work. This material is based upon work supported by the U.S. Department of Energy, Office of Science, National Quantum Information Science Research Centers. Q.W. acknowledges support from the  NSF QuBBE Quantum Leap Challenge Institute (NSF OMA-2121044). The authors thank the Research Computing Center at the University of Chicago for the computing resources.
\end{acknowledgments}

\section*{Supplementary Material}
See the supplementary material (Sec. S1) for both singlet guess and triplet guess results at equilibrium and dissociation. CASCI states at dissociation with singlet guess and triplet guess are listed in Sec. S2. The C-S bonding and antibonding orbital evolution of DPS and TPS dissociation with LASSCF is illustrated in Sec. S3. Sec. S4 lists the absolute energies at equilibrium and dissociation for each system studied with various methods. 
\section*{Author Declarations}
\subsection*{Conflict of Interest}
The authors have no conflicts to disclose.

\subsection*{Author Contributions}
\textbf{Qiaohong Wang}: Data curation (lead); Formal analysis (equal); Investigation (equal); Methodology (equal); Writing – original draft
(lead); Writing – review \& editing (equal). \textbf{Valay Agarawal}: Formal analysis (equal); Investigation (supporting); Methodology (equal); Writing – original
draft (supporting); Writing – review \& editing (equal). \textbf{Matthew R. Hermes}: Conceptualization (equal); Formal analysis (equal); Investigation (equal); Methodology (equal); Writing – original
draft (supporting); Writing – review \& editing (equal).
\textbf{Mario Motta}: Conceptualization (equal); Formal analysis (equal); Investigation (equal); Methodology (equal); Writing – original
draft (supporting); Writing – review \& editing (equal). \textbf{Julia E. Rice}: Conceptualization (equal); Formal analysis (equal); Investigation (equal); Methodology (equal); Writing – original
draft (supporting); Writing – review \& editing (equal).
\textbf{Gavin O. Jones}: Project administration (lead); Conceptualization (equal); Formal analysis (equal); Investigation (equal); Methodology (equal); Writing – original
draft (supporting); Writing – review \& editing (equal).
\textbf{Laura Gagliardi}: Project administration (lead); Resources (lead); Supervision (lead); Conceptualization (equal); Formal analysis (equal); Investigation (equal); Methodology (equal); Writing – original draft (supporting); Writing – review \& editing (equal).

\section*{Data Availability Statement}
The data that support the findings of this study are available
within the article and its supplementary material.

\section*{References}
\bibliography{aipsamp}

\end{document}



\title{Supporting Information: Distinguishing homolytic versus heterolytic bond dissociation of phenyl sulfonium cations with localized active space methods} 

\author{Qiaohong Wang} \affiliation{Pritzker School of Molecular Engineering, University of Chicago, Chicago, IL, 60637, USA.}
\author{Valay Agarawal}
\author{Matthew R. Hermes}
\affiliation{Department of Chemistry, Chicago Center for Theoretical Chemistry, University of Chicago, Chicago, IL 60637, USA.}
\author{Mario Motta}
\affiliation{IBM Quantum, IBM T. J. Watson Research Center, Yorktown Heights, NY 1059}
\author{Julia E. Rice}
\author{Gavin O. Jones}
\affiliation
{IBM Quantum, IBM Research-Almaden, San Jose, CA, 95120, USA.}
\author{Laura Gagliardi}
\email{lgagliardi@uchicago.edu}
\affiliation{Pritzker School of Molecular Engineering, University of Chicago, Chicago, IL, 60637, USA.}
\affiliation{Department of Chemistry, Chicago Center for Theoretical Chemistry, University of Chicago, Chicago, IL 60637, USA.}
\affiliation{James Franck Institute, Chicago Center for Theoretical Chemistry, University of Chicago, Chicago, IL 60637, USA.}
\affiliation{Argonne National Laboratory
9700 S. Cass Avenue
Lemont, IL 60439}
\date{\today}


\date{\today}

\maketitle 
\tableofcontents

\section{Singlet guess and triplet guess result}
\begin{figure}
\centering
\includegraphics[angle=90,origin=c,scale = 0.5]{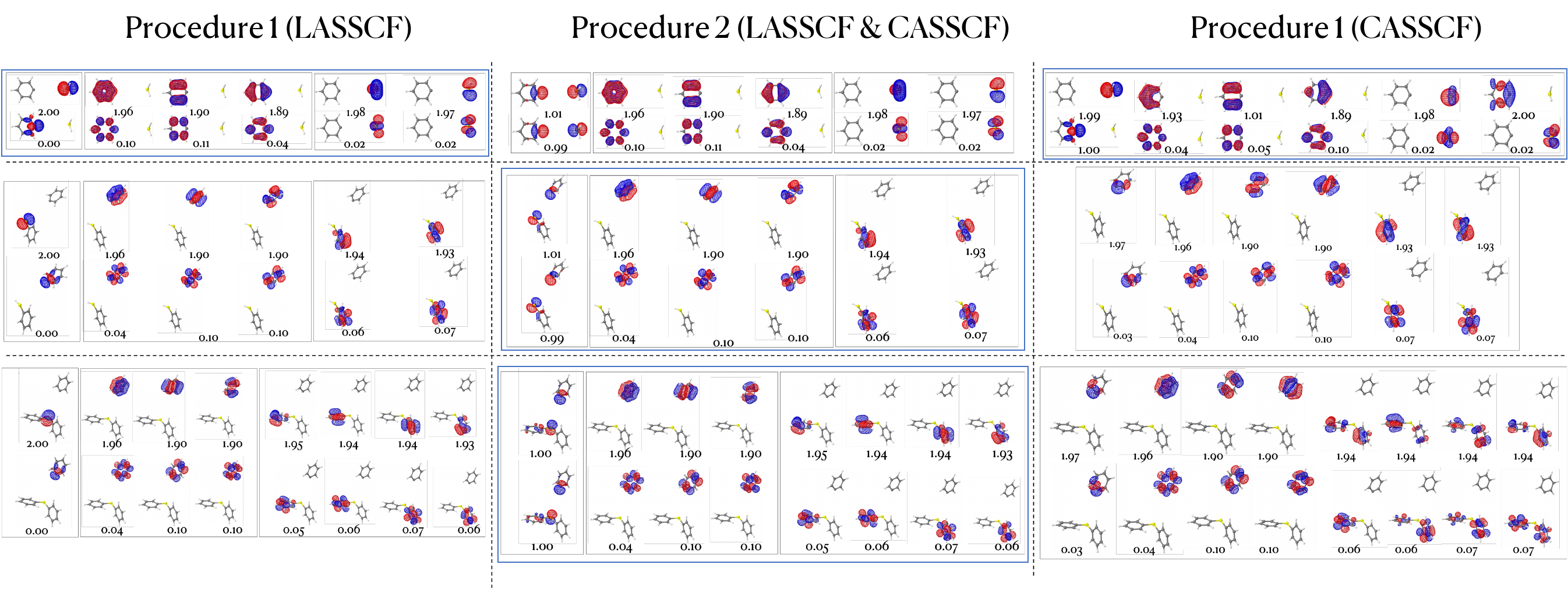}
\caption{}
\label{si_full}
\end{figure}
In Figure \ref{si_full}, we list all the dissociated products with singlet guess (Procedure 1) and triplet guess (Procedure 2) for MPS, DPS, and TPS. Boxed orbitals are the lowest in energy and correspond to the optimized wave function we use the subsequent potential energy scans and dissociation energy calculations.
\clearpage

\section{Confirming ground state solution}
\begin{table}[htbp]
\begin{tabular}{|l|l|l|}
\hline
State & Energy (H.a.) & Characterization                              \\ \hline
S0    & -628.538688   & Heterolytic                                 \\ \hline
S1    & -628.405117   & Heterolytic with $\pi\to p$ excitation \\ \hline
S2    & -628.393863   & Heterolytic with  $\pi\to p$ excitation \\ \hline
S3    & -628.333419   & Homolytic                                   \\ \hline
S4    & -628.319196   & Heterolytic with $\pi\to \pi*$ and $\pi\to p$ excitation  \\ \hline
S5    & -628.299290   & Heterolytic with $\pi\to \pi*$ and $\pi\to p$ excitation  \\ \hline
S6    & -628.215175   & Heterolytic with $\pi\to \pi*$ and $\pi\to p$ excitation   \\ \hline
S7    & -628.209719   & Heterolytic with $\pi\to \pi*$ and $\pi\to p$ excitation   \\ \hline
S8    & -628.207411   & Heterolytic with $\pi\to \pi*$ and $\pi\to p$ excitation   \\ \hline
S9    & -628.200641   & Homolytic                                   \\ \hline
S10   & -628.197235   & Heterolytic with $\pi\to \pi*$, and $\pi\to p$ excitation   \\ \hline
\end{tabular}
\caption{CASCI states at dissociation with singlet guess start for monophenylsulfonium.}
\label{si-monohetero}
\end{table}

\begin{table}
\begin{tabular}{|l|l|l|}
\hline
State & Energy (H.a.) & Characterization                            \\ \hline
S0    & -628.530827   & Homolytic                                   \\ \hline
S1    & -628.493227   & Heterolytic with $\pi\to p$ excitation \\ \hline
S2    & -628.478259   & Heterolytic with $\pi\to p$ excitation \\ \hline
S3    & -628.454218   & Heterolytic with $\pi\to p$ excitation \\ \hline
S4    & -628.406703   & Homolytic                                   \\ \hline
S5    & -628.376541   & Heterolytic with $\pi\to \pi*$ and $\pi\to p$ excitation \\ \hline
S6    & -628.371268   & Homolytic                                   \\ \hline
S7    & -628.357730   & Homolytic                                   \\ \hline
S8    & -628.345105   & Homolytic                                   \\ \hline
S9    & -628.340549   & Heterolytic with $\pi\to \pi*$ and $\pi\to p$ excitation  \\ \hline
S10   & -628.316362   & Homolytic                                   \\ \hline
\end{tabular}
\caption{CASCI states at dissociation with triplet guess start for monophenylsulfonium.}
\label{si-monohomo}
\end{table}

\begin{table}
\begin{tabular}{|l|l|l|}
\hline
State & Energy (H.a.) & Characterization                          \\ \hline
S0    & -858.095781   & Heterolytic                               \\ \hline
S1    & -857.925333   & Heterolytic with $\pi\to p$ excitation    \\ \hline
S2    & -857.922222   & Heterolytic with $\pi\to p$ excitation    \\ \hline
S3    & -857.871751   & Heterolytic with $\pi\to p$ excitation     \\ \hline
S4    & -857.859124   & Heterolytic with $\pi\to p$ excitation     \\ \hline
S5    & -857.851130   & Heterolytic with $\pi\to \pi*$ excitation               \\ \hline
S6    & -857.832670   & Heterolytic with $\pi\to p$ excitation     \\ \hline
S7    & -857.824748   & Heterolytic with $\pi\to p$ excitation     \\ \hline
S8    & -857.791297   & Heterolytic with $\pi\to \pi*$ and $\pi\to p$ excitation \\ \hline
S9    & -857.786425   & Heterolytic with $\pi\to \pi*$ and $\pi\to p$ excitation \\ \hline
S10   & -857.779401   & Heterolytic with $\pi\to \pi*$ and $\pi\to p$ excitation \\ \hline
\end{tabular}
\caption{CASCI states at dissociation with singlet guess start for diphenylsulfonium.}
\label{si-dihetero}
\end{table}

\begin{table}
\begin{tabular}{|l|l|l|}
\hline
State & Energy (H.a.) & Characterization                        \\ \hline
S0    & -858.128602   & Homolytic                               \\ \hline
S1    & -858.031433   & Heterolytic with $\pi\to p$ excitation  \\ \hline
S2    & -858.029699   & Heterolytic with $\pi\to p$ excitation  \\ \hline
S3    & -858.021515   & Heterolytic with $\pi\to p$ excitation   \\ \hline
S4    & -858.009867   & Heterolytic with $\pi\to p$ excitation   \\ \hline
S5    & -857.968645   & Homolytic with $\pi\to \pi*$ excitation              \\ \hline
S6    & -857.956627   & Heterolytic                             \\ \hline
S7    & -857.941540   & Homolytic with $\pi\to \pi*$ excitation                \\ \hline
S8    & -857.939996   & Homolytic with $\pi\to \pi*$ and $\pi\to p$ excitation\\ \hline
S9    & -857.929747   & Homolytic with $\pi\to \pi*$ excitation              \\ \hline
S10   & -857.925525   & Homolytic with $\pi\to \pi*$ excitation               \\ \hline
\end{tabular}
\caption{CASCI states at dissociation with triplet guess start for diphenylsulfonium.}
\label{si-dihomo}
\end{table}

\begin{table}
\begin{tabular}{|l|l|l|}
\hline
State & Energy (H.a.) & Characterization                          \\ \hline
S0    & -1087.695876  & Heterolytic                               \\ \hline
S1    & -1087.521062  & Heterolytic with $\pi\to p$ excitation    \\ \hline
S2    & -1087.520436  & Heterolytic with  $\pi\to \pi*$ excitation    \\ \hline
S3    & -1087.472743  & Heterolytic with $\pi\to p$ excitation     \\ \hline
S4    & -1087.469512  & Heterolytic with $\pi\to p$ excitation     \\ \hline
S5    & -1087.445204  & Heterolytic with $\pi\to \pi*$ excitation               \\ \hline
S6    & -1087.444043  & Heterolytic with $\pi\to p$ excitation                \\ \hline
S7    & -1087.443948  & Heterolytic with $\pi\to \pi*$ excitation              \\ \hline
S8    & -1087.438535  & Heterolytic with $\pi\to \pi*$ excitation and $\pi\to p$\\ \hline
S9    & -1087.427403  & Heterolytic with $\pi\to p$ excitation            \\ \hline
S10   & -1087.418355  & Heterolytic with $\pi\to \pi*$ excitation              \\ \hline
\end{tabular}
\caption{CASCI states at dissociation with singlet guess start for triphenylsulfonium.}
\label{si-trihetero}
\end{table}

\begin{table}
\begin{tabular}{|l|l|l|}
\hline
State & Energy (H.a.) & Characterization                      \\ \hline
S0    & -1087.766521  & Homolytic                             \\ \hline
S1    & -1087.649627  & Homolytic with $\pi\to p$ excitation  \\ \hline
S2    & -1087.639548  & Homolytic with $\pi\to p$ excitation  \\ \hline
S3    & -1087.630845  & Heterolytic with $\pi\to p$ excitation \\ \hline
S4    & -1087.625758  & Heterolytic with $\pi\to p$ excitation \\ \hline
S5    & -1087.615442  & Heterolytic with $\pi\to p$ excitation \\ \hline
S6    & -1087.611609  & Homolytic with $\pi\to p$ excitation            \\ \hline
S7    & -1087.603431  & Heterolytic with $\pi\to p$  excitation          \\ \hline
S8    & -1087.602119  & Homolytic with $\pi\to \pi*$ excitation             \\ \hline
S9    & -1087.591701  & Heterolytic with $\pi\to p$ excitation          \\ \hline
S10   & -1087.589343  & Homolytic with $\pi\to \pi*$ excitation            \\ \hline
\end{tabular}
\caption{CASCI states at dissociation with triplet guess start for triphenylsulfonium.}
\label{si-trihomo}
\end{table}
\clearpage
\section{Homolytic dissociation pathways of DPS and TPS}
\begin{figure}[h!]
\centering
\includegraphics[scale=0.65]{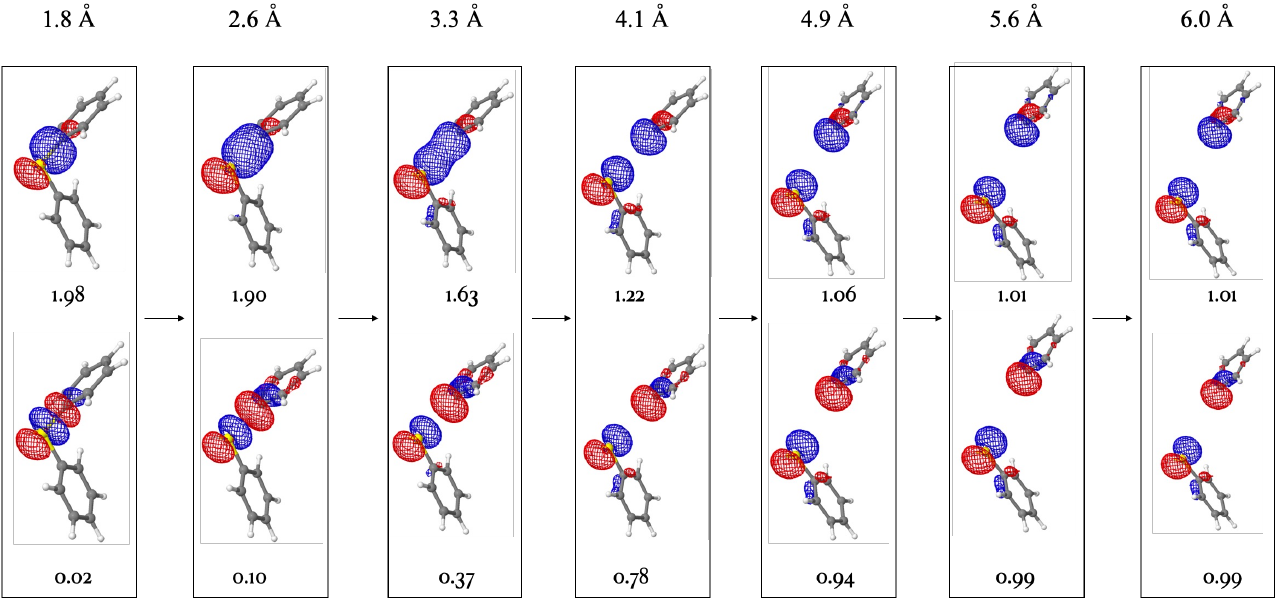}
\caption{DPS ground state, C-S bond homolytic dissociation from LASSCF((2,2),(6,6)) potential energy scan. }
\label{si_dps_orbs}
\end{figure}
\begin{figure}[h!]
\centering
\includegraphics[scale=0.75]{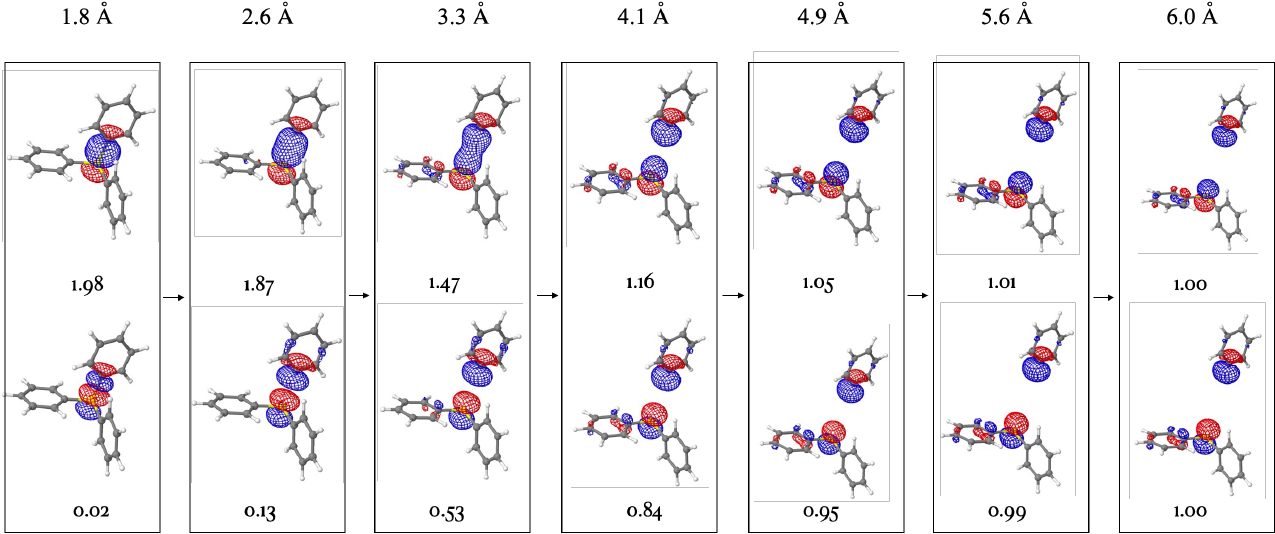}
\caption{TPS ground state, C-S bond homolytic dissociation from LASSCF((2,2),(6,6),(2,2)) potential energy scan. }
\label{si_tps_orbs}
\end{figure}
\clearpage
\section{Absolute energy tables}
\begin{table}[htbp]
\begin{tabular}{llll}
\hline\hline
\multicolumn{1}{c}{\bf{Method}}&
\multicolumn{1}{c}{\bf{Equilibrium  (H.a.)}} &
\multicolumn{1}{c}{\bf{Heterolytic (H.a.)}} &
\multicolumn{1}{c}{\bf{Homolytic (H.a.)}}\\ 
\hline
CASSCF & -628.682175 & -628.556219 & -628.531205  \\
LASSCF & -628.673211& -628.537075 & -628.526406 \\
CASCI\textsuperscript{\emph{a}} &-628.681707  &-628.538688 & -628.530827\\
NEVPT2(CASSCF)&-629.486898 &-629.345975& -629.323871 \\
NEVPT2(CASCI\textsuperscript{\emph{a}})&-629.486919 & -629.340915  & -629.324480\\
LAS-PDFT\textsuperscript{\emph{*}}&-630.218711 &-630.080129&-630.035134\\
CAS-PDFT\textsuperscript{\emph{*}}&-630.206832 & -630.077042 & -630.035786\\
\hline
\end{tabular}

    \textsuperscript{\emph{a}} LAS orbitals;\textsuperscript{\emph{*}} tPBE functional
\caption{Mono-phenyl-sulfonium heterolytic and homolytic absolute energies. }
\label{si-mpstable}
\end{table}

\begin{table}[h!]
\begin{tabular}{ll}
\hline\hline
\multicolumn{1}{c}{\bf{Method}}&
\multicolumn{1}{c}{\bf{Heterolytic (H.a.)}} \\ 
\hline
CASSCF & -628.566198 \\
NEVPT2(CASSCF)& -629.337205\\
CAS-PDFT\textsuperscript{\emph{*}}&-630.077042\\
\hline
\end{tabular}

   \textsuperscript{\emph{*}} tPBE functional
\caption{Mono-phenyl-sulfonium heterolytic energies obtained from forward scan. }
\label{si-mpstable2}
\end{table}

\begin{table}
\begin{tabular}{cllll}
\hline\hline
\multicolumn{1}{c}{\bf{Active Space}}&
\multicolumn{1}{c}{\bf{Method}}&
\multicolumn{1}{c}{\bf{Equilibrium  (H.a.)}} &
\multicolumn{1}{c}{\bf{Heterolytic (H.a.)}} &
\multicolumn{1}{c}{\bf{Homolytic (H.a.)}} \\ 
\hline
\multirow{7}{*}{(8,8)}
&CASSCF& -858.226603 & -858.083205&-858.092514 \\
&LASSCF & -858.221943 &-858.053529&-858.089537  \\   
&CASCI\textsuperscript{\emph{a}} &-858.225928 &-858.054164& -858.092279  \\
&NEVPT2(CASSCF)&-859.859991 & -859.676811&-859.711304   \\
&NEVPT2(CASCI\textsuperscript{\emph{a}}) & -859.859821  &-859.674730&-859.711612  \\
&CAS-PDFT\textsuperscript{\emph{*}}& -860.974076 & -860.818005&-860.818444 \\
&LAS-PDFT\textsuperscript{\emph{*}} &-860.971861  &-860.802549&-860.817947  \\
\hline
\multirow{7}{*}{(10,10)}
&CASSCF&-858.240631 & -858.097184&-858.098101   \\
&LASSCF &-858.236258 & -858.067506 &-858.093433  \\   
&CASCI\textsuperscript{\emph{a}} &-858.239785  &-858.064859&-858.097329  \\
&NEVPT2(CASSCF)&-859.852058 &-859.673964&-859.716986    \\
&NEVPT2(CASCI\textsuperscript{\emph{a}})&-859.856957   &-859.680168& -859.713421  \\
&CAS-PDFT\textsuperscript{\emph{*}}&-860.970681 & -860.796212   & -860.815798  \\
&LAS-PDFT\textsuperscript{\emph{*}} & -860.971858&-860.801090 &-860.804626   \\
\hline
\multirow{7}{*}{(12,12)}
&CASSCF&-858.268657 &-858.124807 &-858.129437 \\
&LASSCF & -858.263417&  -858.095127& -858.124478   \\   
&CASCI\textsuperscript{\emph{a}}& -858.267916&-858.095781 &-858.128602   \\
&NEVPT2(CASSCF)& -859.861176 & -859.678827    & -859.716421   \\
&NEVPT2(CASCI\textsuperscript{\emph{a}}) & -859.860269  &-859.676917   &  -859.711323 \\
&CAS-PDFT\textsuperscript{\emph{*}}&-860.963095 &-860.790250   & -860.808736 \\
&LAS-PDFT\textsuperscript{\emph{*}} &  -860.963009 & -860.793439  & -860.797217 \\
\hline
\end{tabular}

    \textsuperscript{\emph{a}} LAS orbitals;\textsuperscript{\emph{*}} tPBE functional
\caption{Di-phenyl-sulfonium homolytic and heterolytic absolute energy comparison. }
\label{si-dpstable}

\end{table}

\begin{table}
\begin{tabular}{cllll}
\hline\hline
\multicolumn{1}{c}{\bf{Active Space}}&
\multicolumn{1}{c}{\bf{Method}}&
\multicolumn{1}{c}{\bf{Equilibrium  (H.a.)}} &
\multicolumn{1}{c}{\bf{Heterolytic (H.a.)}} &
\multicolumn{1}{c}{\bf{Homolytic (H.a.) }}\\ 
\hline
\multirow{7}{*}{(8,8)}
&CASSCF & -1087.802217 &-1087.642709 &-1087.685344 \\   
&LASSCF &-1087.797461 &-1087.613420&-1087.682440 \\
&CASCI\textsuperscript{\emph{a}} &-1087.801497&  -1087.614028 &-1087.685115   \\
&NEVPT2(CASSCF)&-1090.229321 & -1090.028336&  -1090.096892 \\
&NEVPT2(CASCI\textsuperscript{\emph{a}}) & -1090.229362  & -1090.025452& -1090.097182    \\
&LAS-PDFT\textsuperscript{\emph{*}} &-1091.718939 &-1091.534692 & -1091.586642  \\
&CAS-PDFT\textsuperscript{\emph{*}}&-1091.721117 &-1091.531502 &-1091.587137 \\
\hline
\multirow{7}{*}{(12,12)}&CASSCF&  -1087.830573 &-1087.670181&-1087.714999   \\
&LASSCF& -1087.824494&-1087.640900&-1087.710078\\ 
&CASCI\textsuperscript{\emph{a}}& -1087.828836&-1087.636753& -1087.714312 \\
&NEVPT2(CASSCF) &-1090.214026 &-1090.019017&-1090.092672  \\
&NEVPT2(CASCI\textsuperscript{\emph{a}})&-1090.224187  &-1090.032173&-1090.091116    \\
&CAS-PDFT\textsuperscript{\emph{*}}&-1091.711594  &-1091.519610&-1091.584237  \\
&LAS-PDFT\textsuperscript{\emph{*}} &-1091.714261  &-1091.530486&-1091.581262 \\
\hline
\multirow{7}{*}{(16,16)}&CASSCF& -1087.884623  &-1087.724467&-1087.768893 \\
&LASSCF & -1087.878777&-1087.695112& -1087.758508  \\
&CASCI\textsuperscript{\emph{a}}&-1087.883808&-1087.695876&-1087.766521 \\
&NEVPT2(CASSCF)& -1090.229376 & -1090.031554&-1090.101492  \\
&NEVPT2(CASCI\textsuperscript{\emph{a}}) &-1090.230884 &-1090.029187 &-1090.096098  \\
&CAS-PDFT\textsuperscript{\emph{*}}&-1091.704832 &-1091.522568&-1091.572549  \\
&LAS-PDFT\textsuperscript{\emph{*}} &-1091.701301 &-1091.516701&-1091.561975 \\
\hline
\end{tabular}

    \textsuperscript{\emph{a}} LAS orbitals;\textsuperscript{\emph{*}} tPBE functional
\caption{Tri-phenyl-sulfonium heterolytic and homolytic absolute energy comparison.  }
\label{SI-tpstable}
\end{table}

\begin{table}
\begin{tabular}{clccc}
\hline\hline
\multicolumn{1}{c}{\bf{Molecule}}&
\multicolumn{1}{c}{\bf{Functional}}&
\multicolumn{1}{l}{\bf{Equilibrium  (H.a.)}} &
\multicolumn{1}{l}{\bf{Dissociation (H.a.)}} &
\multicolumn{1}{c}{\bf{Spin density and Character}} \\ 
\hline
\multirow{5}{*}{MPS}
&B3LYP-d3bj&-630.889017 & -630.769482 &0.355128, N.A. \\
&TPSSh-d3bj&-630.898309 &-630.782906 & 0.369196, N.A.  \\ 
&BP86-d3bj&-630.905188 & -630.797168 &0.380381, N.A.\\
&M062X&-630.713571 & -630.574032 & 0.245793, heterolytic\\
&MN15&-630.428849 & -630.294249 & 0.000000, heterolytic\\
\hline
\multirow{5}{*}{DPS}
&B3LYP-d3bj&-862.072484&-861.941091&0.640340, N.A.\\
&BP86-d3bj&-862.082008&-861.960539&0.591855, N.A. \\
&TPSSh-d3bj&-862.090584 &-861.956429&0.626163, N.A. \\
&M062X&-861.773649 &-861.623629&0.760625, homolytic\\
&MN15&-861.294204 &-861.150632 &0.646896, N.A.\\
\hline
\multirow{5}{*}{TPS}
&B3LYP-d3bj &-1093.253393 & -1093.124935& 0.819654, homolytic \\
&TPSSh-d3bj& -1093.279624  &-1093.155324&0.783681, homolytic \\
&BP86-d3bj & -1093.254191&-1093.131802&0.705203, homolytic  \\
&M062X &-1092.831797  &  -1092.694104& 0.999084, homolytic \\
&MN15 &-1092.156142 & -1092.018473 &0.912824, homolytic\\
\hline

\end{tabular}

\caption{DFT results for MPS, DPS and TPS. }
\label{si-dfttable}

\end{table}


%
%

%

